\documentstyle[12pt]{article}
\begin{document}
\thispagestyle{empty}
\noindent
\hspace*{\fill} KA--THEP--1--1995 \\
\hspace*{\fill} UAB--FT--357 \\
\hspace*{\fill} January 1995 \\
\vspace{1.5cm}
\begin{center}
\begin{large}
\begin{bf}
STRONG SUPERSYMMETRIC QUANTUM EFFECTS ON THE TOP QUARK WIDTH\\
\end{bf}
\end{large}
\vspace{1cm}
Andreas DABELSTEIN, Wolfgang HOLLIK, Christoph J\"UNGER\\

\vspace{0.25cm}
Institut f\"ur Theoretische Physik,\\
 Universit\"at Karlsruhe,
D-76128 Karlsruhe, Germany\\

\vspace{0.25cm}
Ricardo A. JIM\'ENEZ, Joan SOL\`A\\

\vspace{0.25cm}
Grup de F\'{\i}sica Te\`orica\\
and\\
Institut de F\'\i sica d'Altes Energies\\
\vspace{0.25cm}
Universitat Aut\`onoma de Barcelona\\
08193 Bellaterra (Barcelona), Catalonia, Spain\\
\end{center}
\vspace{0.3cm}
\begin{center}
{\bf ABSTRACT}
\end{center}
\begin{quotation}
\noindent
We compute the one-loop supersymmetric QCD
quantum effects on the width
$\Gamma (t\rightarrow W^{+}\, b)$
of the canonical main decay of the top quark within the framework of the
MSSM. The corrections can be of either sign depending on whether the
stop squark mass is above or below the top quark decay threshold into
stop and gluino $\Gamma (t\rightarrow\tilde{t}\,\tilde{g})$.
For $m_{\tilde{t}}$ above that threshold, the corrections are negative
and can be of the same order (and go in the same direction) as the
ordinary QCD corrections, even for stop and gluino masses of
${\cal O}(100)\,GeV$. Since the electroweak
supersymmetric quantum effects turn out to be also of the same sign and could
be of the same order of magnitude, the total MSSM correction to the  top quark
width could potentially result in a rather large ${\cal O}(10-25)\%$ reduction
 of
$\Gamma (t\rightarrow W^{+}\, b)$ far beyond the conventional QCD expectations.
\end{quotation}

\baselineskip=6.5mm  

\newpage

{\bf 1. Introduction}\\[0.3cm]

In spite of being a sequential fermion the top quark plays a special
role in the families of fermions.
The Standard Model predicts a fairly strong direct interaction with
the Higgs sector through a large Yukawa coupling distinguishing the
top quark as a particularly helpful probe for the electroweak
symmetry breaking mechanism. It may also be a useful laboratory
to unravel effects beyond the Standard Model
such as those from the minimal supersymmetric standard model (MSSM)
\cite{b005}.
For realizing possible hints for new physics of this class of
extensions in future precise measurements of the
top quark properties detailed predictions are necessary including
higher order terms.
The quantum effects to the conventional top width
$\Gamma(t\rightarrow W^+ b)$
induced by the one-loop radiative corrections
from the
chargino, neutralino and scalar fermion sector of the MSSM have been studied
in detail in \cite{b001} (see also \cite{b002}) with the main result that there
can be relatively large corrections from this ``genuine'' electroweak
SUSY part of the MSSM (up to 10\%) depending sensitively
on the range of
the model parameters.
These potentially large contributions are of the same (negative) sign
as the standard QCD corrections \cite{b111,b211} and of the same order of
magnitude.
This is different from the standard electroweak corrections which have
been shown to be small with very little variation on the mass of the top
quark and the Higgs boson \cite{b211,b013,b213}.
Also the non-standard loop effects from the SUSY Higgs sector
are well below 1\% \cite{b014,b022}.

\smallskip
In this paper we complete the discussion of the one-loop quantum effects
on the top quark width in the framework of the MSSM by computing the
missing strong SUSY corrections induced by  virtual gluinos
and scalar quarks.
Due to the strength of the strong coupling constant these additional loop
effects are expected to be comparable with the electroweak terms arising
from large Yukawa couplings if the gluino and squark masses are not too
high. Our study shows that the most interesting virtual effects (i.e.
when the direct real decay channel $t \rightarrow \tilde{t} \tilde{g}$
is kinematically forbidden) have a similar signature:
the SUSY QCD contributions are negative and can reach several percent,
thus enhancing the conventional QCD corrections nearly by a factor
of two. \\[0.3cm]

{\bf 2. Supersymmetric QCD corrections}\\[0.3cm]

To compute the one-loop QCD corrections to
 $\Gamma_t\equiv\Gamma (t\rightarrow W^{+}\, b)$ in the MSSM, we shall adopt
the on-shell renormalization scheme, where the fine structure constant,
$\alpha$, and the masses of the gauge bosons, fermions and scalars are
the renormalized parameters: $(\alpha, M_W, M_Z, M_H, m_f, M_{SUSY},...)$
\footnote{ For a comprehensive review, see e.g. refs. \cite{b020,b021}.}.
There are no universal supersymmetric corrections from the strong
interaction at 1-loop order. The flavor specific vertex corrections and
quark self-energies originating from virtual gluinos and squarks (stop and
sbottom species) are depicted in Fig.1.
The SUSY-QCD interaction Lagrangian relevant to our calculation is given,
in four-component notation, by
\begin{equation}
{\cal L}= -{g_s\over\sqrt{2}}\,\left[\tilde{q}^{i *}_{L}\,(\lambda_r)_{ij}\,
\bar{\tilde{g^r}}\,P_L\,q^j-\bar{q}^i(\lambda_r)_{ij}\,
P_L\,{\tilde{g^r}}\,\tilde{q}^{j}_{R}\right]+{\rm h.c.}\,,
\end{equation}
where $\tilde{g}^r (r=1,2,...,8)$ are the Majorana gluino fields,
$(\lambda_r)_{ij} (i,j=1,2,3) $ are the Gell-Mann matrices, and
$\tilde{q'}_a=\{\tilde{q}_L,\tilde{q}_R\}$ are the weak-eigenstate squarks
associated to the two chiral components  $P_{L,R}\,q=1/2(1\mp\gamma_5)\,q$;
they are related to
the corresponding mass-eigenstates $\tilde{q}_a$
by a rotation $2\times 2$ matrix (we neglect intergenerational mixing):
\begin{eqnarray}
\tilde{q}_a&=&\sum_{b} R_{ab}^{(q)}\tilde{q'}_b,\nonumber\\
R^{(q)}& =&\left(\begin{array}{cc}
\cos{\theta_q}  &  \sin{\theta_q} \\
-\sin{\theta_q} & \cos{\theta_q}
\end{array} \right)\;\;\;\;\;\;
(q=t, b)
\label{eq:rotation}
\end{eqnarray}
Using minimal field renormalization \cite{b020} (viz. one renormalization
 constant
per symmetry multiplet), the renormalized vertex can be written
in the notation of Ref.\,\cite{b001} as follows:
\begin{equation}
\Gamma_{\mu}^{(1)}=i\,{g\over \sqrt{2}}
[\gamma_{\mu}\,P_L(1+F_L+\delta Z_L-{1\over 2}\hat\Pi_t(m_t^2))+
\gamma_{\mu}\,P_R\,F_R+{p_{\mu}\over M_W}(P_L\,H_L+P_R\,H_R)]\,,
\label {eq:RENV}
\end{equation}
where the correction has been parametrized in terms of four form factors
$F_L, F_R, H_L$ and $ H_R$.
The renormalized one-loop vertex is obtained from ${\cal L}\rightarrow
{\cal L}+\delta{\cal L}$ , where $\delta{\cal L}$ is calculated from
$Z_L$ being the renormalization constant of the $(t,b)$ doublet:
\begin{equation}
\left(\begin{array}{c}t\\b\end{array}\right)\rightarrow Z_L^{1/2}
\left(\begin{array}{c}t\\b\end{array}\right).
\end{equation}
Fixing $Z_L=1+\delta Z_L$ through the condition of residue one for the
renormalized bottom quark propagator
\begin{equation}
 \frac{1}{\not{k}-m_b}\hat{\Sigma}^b\left(k\right) u_b\left(k\right)
 \bigg|_{\not{k}\rightarrow m_b}=0\,,
\end{equation}
yields for $\delta Z_L$
\begin{equation}
 \delta Z_L=\Sigma^b_L(m_b^2)+m_b^2\left[
 {\Sigma^b_L}'(m_b^2)+{\Sigma^b_R}'(m_b^2)+
 2\:{\Sigma^b_S}'(m_b^2)\right]
\end{equation}
and a finite wave-function renormalization for the external top quark line
(in the framework of \cite{b020}):
\begin{equation}
 \hat{\Pi}_t(m_t^2)=\Pi_t(m_t^2)+\delta Z_L\,,
\end{equation}
with $\Pi_t(m_t^2)$ given by
\begin{equation}
\Pi_t(m_t^2)=-\Sigma^t_L(m_t^2)-m_t^2\left[{\Sigma^t_L}'(m_t^2)
 +{\Sigma^t_R}'(m_t^2)+2\;{\Sigma^t_S}'(m_t^2)\right].
\end{equation}
$\delta Z_L$ and $\hat{\Pi}_t$ are obtained from the self energy diagrams
of Fig.1, for which we used the decomposition
\begin{equation}
 \Sigma^q(p)=\Sigma^q_L(p^2) \not{p} P_L +
               \Sigma^q_R(p^2) \not{p} P_R +
            m_q \Sigma^q_S(p^2)
\end{equation}
and the notation $\Sigma'(p)\equiv \partial\Sigma(p)/\partial p^2$.
Explicitly, the gluino-squark contributions read
\begin{eqnarray}
 \Sigma^q_L(p^2)&=&\frac{\alpha_s}{3\pi}\left[(1+\cos 2\theta_q)B_1(
  p^2;m_{\tilde{g}},m_{\tilde{q}_1})+(1-\cos 2\theta_q)B_1(
  p^2;m_{\tilde{g}},m_{\tilde{q}_2})\right]
  \nonumber\\
 \Sigma^q_R(p^2)&=&\frac{\alpha_s}{3\pi}\left[(1-\cos 2\theta_q)B_1(
  p^2;m_{\tilde{g}},m_{\tilde{q}_1})+(1+\cos 2\theta_q)B_1(
  p^2;m_{\tilde{g}},m_{\tilde{q}_2})\right]
\nonumber\\
\Sigma^q_S(p^2)&=&\frac{\alpha_s}{3\pi}\left[\frac{m_{\tilde{g}}}{m_q}
  \sin 2\theta_q \left(B_0(p^2;m_{\tilde{g}},m_{\tilde{q}_1})-
         B_0(p^2;m_{\tilde{g}},m_{\tilde{q}_2})\right)\right]\,,
\end{eqnarray}
$\theta_q$ being the left-right mixing angle defined in (\ref{eq:rotation})
which diagonalizes
the $\tilde{q}$-squark mass matrix  ${\cal M}_{\tilde{q}}^2$
( $\tilde{q}= \tilde{t}, \tilde{b})$.
As for the vertex diagram of Fig.1, the generic contribution to
the form factors in eq.(\ref{eq:RENV}) is given by
\begin{eqnarray}
F_L&=&\frac{\alpha_s}{\pi} C_F\: a_3 a_{1L} a_{2R}\: C_{00}
 \nonumber \\
F_R&=&\frac{\alpha_s}{\pi} C_F\: a_3 a_{1R} a_{2L}\: C_{00}
  \nonumber\\
H_L&=&\frac{\alpha_s}{\pi} C_F\: a_3 M_W \left[ -a_{1L} a_{2L}
       \:m_{\tilde{g}}\:(C_0+C_1+C_2)
   \right. \nonumber\\ & & \left.
                     +\: a_{1R} a_{2L} \:m_t\: (C_1+C_{11}+C_{12})
                     +   a_{1L} a_{2R} \:m_b\: (C_2+C_{12}+C_{22}) \right]
   \nonumber\\
H_R&=&\frac{\alpha_s}{\pi} C_F\: a_3 M_W \left[ -a_{1R} a_{2R}
       \:m_{\tilde{g}}\:(C_0+C_1+C_2)
   \right. \nonumber\\ & & \left.
                     +\: a_{1L} a_{2R}\:m_t\: (C_1+C_{11}+C_{12})
                     +   a_{1R} a_{2L}\:m_b\: (C_2+C_{12}+C_{22}) \right].
\label{eq:FR}
\end{eqnarray}
where the colour factor $C_F$ is $4/3$. The notations and conventions for
$B_0, B_1$ and the various three-point functions
$C_{\ldots}=C_{\ldots}\big(p^2,(p-p')^2,p'^2,m_{\tilde{g}},m_1,m_2)$
were adopted from \cite{b021}.
There are four contributions of the type (\ref{eq:FR}), which we shall
denote with the indices
$(k,l)=(1,1),(1,2),(2,1), (2,2)$. The corresponding masses and
couplings associated to these contributions can be written in a compact form
as follows:
\begin{eqnarray}
m_1 &=& m_{\tilde{t}_{k}},\ \  \ \  \ m_2=m_{\tilde{b}_{l}},\nonumber\\
a_3 &=& (-1)^{k+l}\sin\left(\theta_t-k\frac{\pi}{2}\right)\,
\sin\left(\theta_b-l\frac{\pi}{2}\right),\nonumber\\
a_{1L} &=&(-1)^{k+1}\sin\left(\theta_t-k\frac{\pi}{2}\right),\nonumber\\
a_{2L} &=&(-1)^{l+1}\cos\left(\theta_b-l\frac{\pi}{2}\right),\nonumber\\
a_{1R} &=&(-1)^{k+1}\cos\left(\theta_t-k\frac{\pi}{2}\right),\nonumber\\
a_{2R} &=&(-1)^{l+1}\sin\left(\theta_b-l\frac{\pi}{2}\right)\,.
\end{eqnarray}
To obtain the result for $\Gamma(t\rightarrow b W^+)$ we define, in analogy
to Ref.\cite{b001}, the four standard matrix elements
\begin{eqnarray}
M_0&=&\bar{u}_b(p') \not{\epsilon} P_L u_t(p)
  \nonumber\\
M_1&=&\bar{u}_b(p') \not{\epsilon} P_R u_t(p)
  \nonumber\\
M_2&=&\bar{u}_b(p') P_L u_t(p) \:(\epsilon.p)\:/M_W
   \nonumber\\
M_3&=&\bar{u}_b(p') P_R u_t(p) \:(\epsilon.p)\:/M_W.
\end{eqnarray}
Then the one loop width can be written in terms of the Born width $\Gamma_0$
and
\begin{eqnarray}
G_0&=&\Sigma_{pol}\left|M_0\right|^2 = m_t^2+m_b^2-2\:M_W^2+
     \frac{(m_t^2-m_b^2)^2}{M_W^2},
  \nonumber \\
G_1&=&\Sigma_{pol}M_0 M_1^{*}=-6\:m_t\;m_b,
   \nonumber\\
G_2&=&\Sigma_{pol}M_0 M_2^{*}=-\frac{m_t}{M_W}\left[ m_t^2+m_b^2 -
     \frac{M_W^2}{2}-
     \frac{(m_t^2-m_b^2)^2}{2\;M_W^2} \right],
   \nonumber\\
G_3&=&\Sigma_{pol}M_0 M_3^{*}=\frac{m_b}{m_t}\: G_2\,,
\end{eqnarray}
and gives
\begin{eqnarray}
\Gamma&=&\Gamma_0 \left( 1+2\,Re F_L + 2 Re\,\delta Z_L-
  Re\,\hat{\Pi}_t (m_t^2))  \right.\nonumber\\
       & &\left.  +2\frac{G_1}{G_0}\,Re\, F_R+
                  2\frac{G_2}{G_0}\,Re H_L+
                  2\frac{G_3}{G_0}\,Re H_R\right)\,,
\end{eqnarray}
with \footnote{$\lambda(x,y,z)=x^2+y^2+z^2-2\;(x y+x z+y z)$}
\begin{eqnarray}
\Gamma_0&=&\left({G_F M_W^2\over 8\pi\sqrt{2}}\right)\,m_t\,|V_{tb}|^2
{G_0\,\lambda^{1/2} (m_t^2 ,M_W^2 ,m_b^2)\over m_t^4}\nonumber\\
&\simeq &\left({G_F m_t^3\over 8\pi\sqrt{2}}\right)\,
\left(1-{M_W^2\over m_t^2}\right)^2\,\left(1+2{M_W^2\over m_t^2}\right)\,.
\label{eq:treeGF}
\end{eqnarray}
A correction term from $\Delta r$ does not appear due to the absence of
1-loop SUSY QCD corrections in $\mu$-decay.\\[0.3cm]

{\bf 3. Numerical analysis}\\[0.3cm]

We are now in position to analyze the relative correction
\begin{equation}
\delta_{\tilde{g}}={\Gamma-\Gamma_0\over \Gamma_0}
\end{equation}
induced by the strong virtual gluino effects.
The corresponding supersymmetric electroweak corrections induced by charginos,
neutralinos and sfermions have already been
addressed in Refs.\cite{b001,b002}, and will be briefly commented later on.
The numerical analysis of the strong supersymmetric
corrections is displayed in Figs.2-6. We used
\begin{equation}
\alpha_s(m_t)=0.11
\end{equation}
as an input for the strong coupling constant, which
remains essentially unchanged within the CDF range
for the top quark mass: $160\,GeV\leq m_t\leq 190\,GeV$\,\cite{D001}.
Whenever a fixed value for $m_t$ is chosen, we take the central CDF
value $m_t=174\,GeV$.
In Figs.2a-2b we plot $\delta_{\tilde{g}}$ as a function of
$m_{\tilde{g}}$ and of
$m_t$, respectively. For the squark mass spectrum we have
borrowed the standard pattern expected from models with radiatively induced
breaking of the electroweak symmetry such as supergravity inspired
models\,\cite{b005}:
\begin{equation}
m^2_{{\tilde{q}}_{L,R}}=m^2_q+M^2_{{\tilde{q}}_{L,R}}\pm
\cos{2\beta}\,(T^3_{L,R}-Q_{\tilde{q}}\,\,s_w^2)\,M_Z^2\,,
\label{eq:sferm}
\end{equation}
where $T^3_{L,R}$ and $Q_{\tilde{q}}$ stand, respectively,
for the third component of weak isospin and electric charge
corresponding to each member of the multiplet and for each ``chiral'' species
$\tilde{q}_{L,R}$ of squarks. In that equation $\tan\beta$ stands for the
ratio $v_2/v_1$ of vacuum expectation values giving masses to the
 $T^3=+1/2$ ($v_2$) and the $T^3=-1/2$ ($v_1$) components in each fermion
doublet. Finally, the parameters
$M_{{\tilde{q}}_{L,R}}$ are soft SUSY-breaking mass terms\,\cite{b005}.
The mass splitting between the $T^3=+1/2$ and the $T^3=-1/2$ components
is independent of $M_{{\tilde{q}}_L}$ because of $SU(2)_L$-gauge invariance,
which requires $M_{\tilde{t}_L}=M_{\tilde{b}_L}$.
For the $(t,b)$ doublet we have
\begin{equation}
m^2_{\tilde{t}_L}
-m^2_{\tilde{b}_L}= m_t^2 +M_W^2\,\cos{2\beta}\,,
\label{eq:splitting1}
\end{equation}
where the bottom quark mass is neglected.
The previous equations refer to truly mass eigenvalues for the squarks only in
the absence of mixing. Thus, in Fig.2 we assumed zero mixing for both stop and
sbottom mass matrices. The effect of the mixing will be examined in a
second step. Furthermore,
since $M_{{\tilde{q}}_R}$ is independent of  $M_{{\tilde{q}}_L}$, we may
assume for simplicity that the $R$-type and $L$-type species are degenerate
in mass. (This assumption will be dropped for stop squarks in the presence
of mixing).
In Fig.3, on the other hand, we test the sensitivity of the SUSY-QCD
corrections
on $\tan\beta$, which is very weak since this parameter enters only
through the squark mass formula (\ref{eq:sferm}). Thus we shall take
 $\tan\beta=1$
for the other plots.
This is in contrast to the
electroweak SUSY corrections, which are very sensitive to $\tan\beta$,
for it may induce large supersymmetric Yukawa couplings\,\cite{b001}.
{}From Figs.2-3, it is clear that in the absence of mixing the
strong SUSY corrections are typically of the order of $-1\%$.
For light gluinos of ${\cal O}(1)\,GeV$\,\cite{b043}, the correction
is $2-3$ times higher (in absolute value) than for
${\cal O}(100)\,GeV$ gluinos,
but it never goes beyond $-2\%$ for sbottom masses
$m_{\tilde b}\stackrel{\scriptstyle >}{{ }_{\sim}} 70\,GeV$. This is because
the relation (\ref{eq:splitting1}) forces the stop squarks to be rather heavy
(${\cal O}(200)\,GeV$).
 Only for gluinos of ${\cal O}(1)\,GeV$ and $m_{\tilde{b}}=45\,GeV$
(the strict LEP limit on squarks) one can scarcely
 reach $\delta_{\tilde{g}}\simeq -3\%$.
As in Ref.\cite{b001}, we have explicitly checked (Cf. Fig.4) that for our
choices of the squark masses, the induced
deviations of the $\rho$-parameter from 1 satisfy
\begin{equation}
|\delta\rho|\leq 0.005\,.
\label{eq:rho}
\end{equation}

The effect of the mixing is studied in Figs. 5-6. To illustrate the
potentiality of this effect, it will suffice to concentrate on the stop mass
 matrix,
where it is most likely to arise:
\begin{equation}
{\cal M}_{\tilde{t}}^2 =\left(\begin{array}{cc}
M_{\tilde{b}_L}^2+m_t^2+\cos{2\beta}({1\over 2}-
{2\over 3}\,s_w^2)\,M_Z^2
 &  m_t\, M_{LR}\\
m_t\, M_{LR} &
M_{\tilde{t}_R}^2+m_t^2+{2\over 3}\,\cos{2\beta}\,s_w^2\,M_Z^2\,.
\end{array} \right)\,.
\label{eq:stopmatrix}
\end{equation}
For $M_{LR}=0$ one recovers eq.(\ref{eq:sferm}), but for nonvanishing $M_{LR}$
a light mass eigenvalue is possible. As a matter of fact, a light stop is
still phenomenologically allowed\,\cite{b1STE}, even below the
purported lower LEP limit of $45\, GeV$ on all
charged particles\,\cite{b1SJA}.
In general, once $\tan\beta$ and $m_{\tilde{b}}$ are
fixed, we are free to choose two independent parameters in the stop mass
matrix:
$(M_{\tilde{t}_R}, M_{LR})$, which, if desired, can be conveniently traded for
$(m_{\tilde{t}_1},\theta_t)$, $(m_{\tilde{t}_1},M_{LR})$ etc.,
$m_{\tilde{t}_1}$ being the lightest of the two mass eigenvalues.
As for the mixing parameter $M_{LR}$ it is in principle arbitrary, with
the caveat that
\begin{equation}
M_{LR}\leq 3\,m_{\tilde{b}_L}\,.
\label{eq:MLR}
\end{equation}
It roughly corresponds to a well-known
necessary, though not sufficient, condition to avoid colour-breaking vacua
\,\cite{b134}.
In Fig.5 we display the dependence of $\delta_{\tilde{g}}$ on $M_{LR}$ for
fixed $m_{\tilde{b}}=80\,GeV$, $\theta_t=\pi/4$ and various gluino masses.
We see that $\delta_{\tilde{g}}$ can be of either sign and reach the $\pm 2\%$
level in the case of light gluinos.
However, for large enough $M_{LR}$ (i.e. small enough $m_{\tilde{t_1}}$)
we can approach regions where a few percent correction is also possible
for ${\cal O}(100)\,GeV$ gluinos.
Finally, in Fig.6 we display the optimal situation where the strong SUSY
corrections to the top quark width are the largest. Here
we have fixed the rather conservative values
 $m_{\tilde{b}}=m_{\tilde{g}}=120\,GeV$
for the sbottom and gluino masses (compatible with the ``traditional''
CDF bounds\,\cite{b044}). In contrast to Fig.5, where $\theta_t$ was kept fixed
at a specific value, we now
let it vary by computing contour lines of constant $\delta_{\tilde{g}}$ in
the $(M_{LR},m_{\tilde{t}_1})$-plane. In these conditions, there is a threshold
(pseudo) singularity\,\cite{D004}
(similar to the one found in Fig.5)
associated to the wave-function renormalization of the
top quark line at $m_{\tilde{t}_1}=54\,GeV$. We cannot arbitrarily approach
from above the (dashed) threshold line in Fig.6
without breaking perturbation theory.
Most remarkably, however, even staying prudentially away from
it we can obtain ${\cal O}(-10)\%$ corrections. On the contrary, if we
approach the threshold line from below (a non-singular limit),
the correction is positive and of order $5\%$.
Nevertheless, it should be clear that the most interesting scenario for our
decay corresponds to $\delta_{\tilde{g}}<0$, in which case the
two-body supersymmetric decay $t\rightarrow \tilde{{t}_1}\tilde{g}$ is
phase-space blocked up. For this situation the strong supersymmetric
corrections could be reinforced by the additional negative
contributions from the electroweak supersymmetric
sector of the MSSM\,\cite{b001}.

Some words on previous work are in order. In Ref.\cite{D002}, a first study
of $\delta_{\tilde{g}}$ as a function of $m_t$ and $m_{\tilde{g}}$ was
 presented.
We qualitatively confirm their results. In that reference, however,
the impact from the mixing and threshold effects
were completely missed and only the simplest situation,
characterized by degenerate masses,
was considered\,\footnote{Notice that the assumption of stop masses equal
to sbottom masses is incompatible with eq.(\ref{eq:splitting1}).}.
 Even in this case, i.e. adapting ourselves to the set of hypotheses
and exact inputs considered in that reference, we
found a relative deviation of $30-40\%$ (our result being higher than theirs)\,
\footnote{We carefully checked this discrepancy by performing two completely
independent computations of $\delta_{\tilde{g}}$ and got perfect agreement with
 the alternative results presented here.}.\\[0.3cm]

{\bf 4. Conclusions}\\[0.3cm]

In summary, the strong supersymmetric corrections to the standard top quark
decay mode $t\rightarrow W^{+}\, b$ are potentially large:
to wit, of order $+ 5\%$ to $-(5-10)\%$.
Remarkably enough, the most significant signature appears for sufficiently
heavy gluinos (compatible with conservative CDF limits) and
in the presence of an intermediately heavy stop squark $\tilde{t}_1$
with a mass slightly beyond the threshold for the direct supersymmetric
decay $t\rightarrow \tilde{{t}_1}\tilde{g}$, i.e. for
$m_{\tilde{t}_1}={\cal O}(50-60)\,GeV$.
 However, even for $m_{\tilde{t}_1}={\cal O}(80-100)\,GeV$ the correction could
remain of order $-5\%$. The case $\delta_{\tilde{g}}<0$ is specially
significant
in that it would add up to the conventional QCD corrections
($\delta_{QCD}\simeq -8\%$) and in favourable circumstances the total strong
correction could reach $-(15-20)\%$. Last but not least, as
this correction is insensitive to $\tan\beta$, we may envisage a situation with
large $\tan\beta\geq m_t/m_b$ in which the electroweak supersymmetric
corrections\,\cite{b001},
being also negative, are of the same order of magnitude as the ones
studied here, in which case the total MSSM correction to the top quark width
(the Higgs correction being negligible\,\cite{b014}) could result in a
remarkable reduction
of $\Gamma (t\rightarrow W^{+}\, b)$ by about $25\%$.

\vspace{0.5cm}

{\bf Acknowledgements}:

The work of RJ and JS has  been partially supported by CICYT
under project No. AEN93-0474.

\newpage
\begin{center}
\begin{Large}
{\bf Figure Captions}
\end{Large}
\end{center}
\begin{itemize}
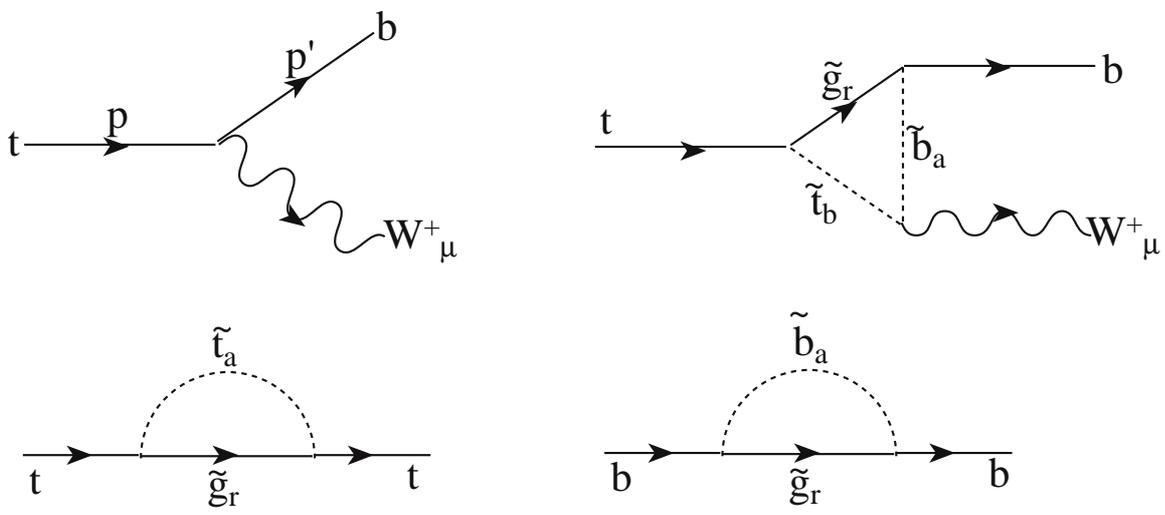
\item{\bf Fig.1} Feynman diagrams, up to one-loop order, for the SUSY-QCD
corrections to the top quark decay $t\rightarrow W^{+}\,b$.
Each one-loop diagram is summed over the mass-eigenstates of the
stop and sbottom squarks
($\tilde{b}_a, \tilde{t}_b\,; a,b=1,2$).

\item{\bf Fig.2} Dependence of $\delta_{\tilde{g}}$ on (a) $m_{\tilde{g}}$
for $m_t=174\,GeV$ and a wide range of SUSY masses including the
light gluino region,
and on (b)  $m_t$ (within the CDF limits) for the same squark
masses as in (a) and $m_{\tilde{g}}=m_{\tilde{b}}$.
In both cases we assume no squark mixing and $\tan\beta=1$.

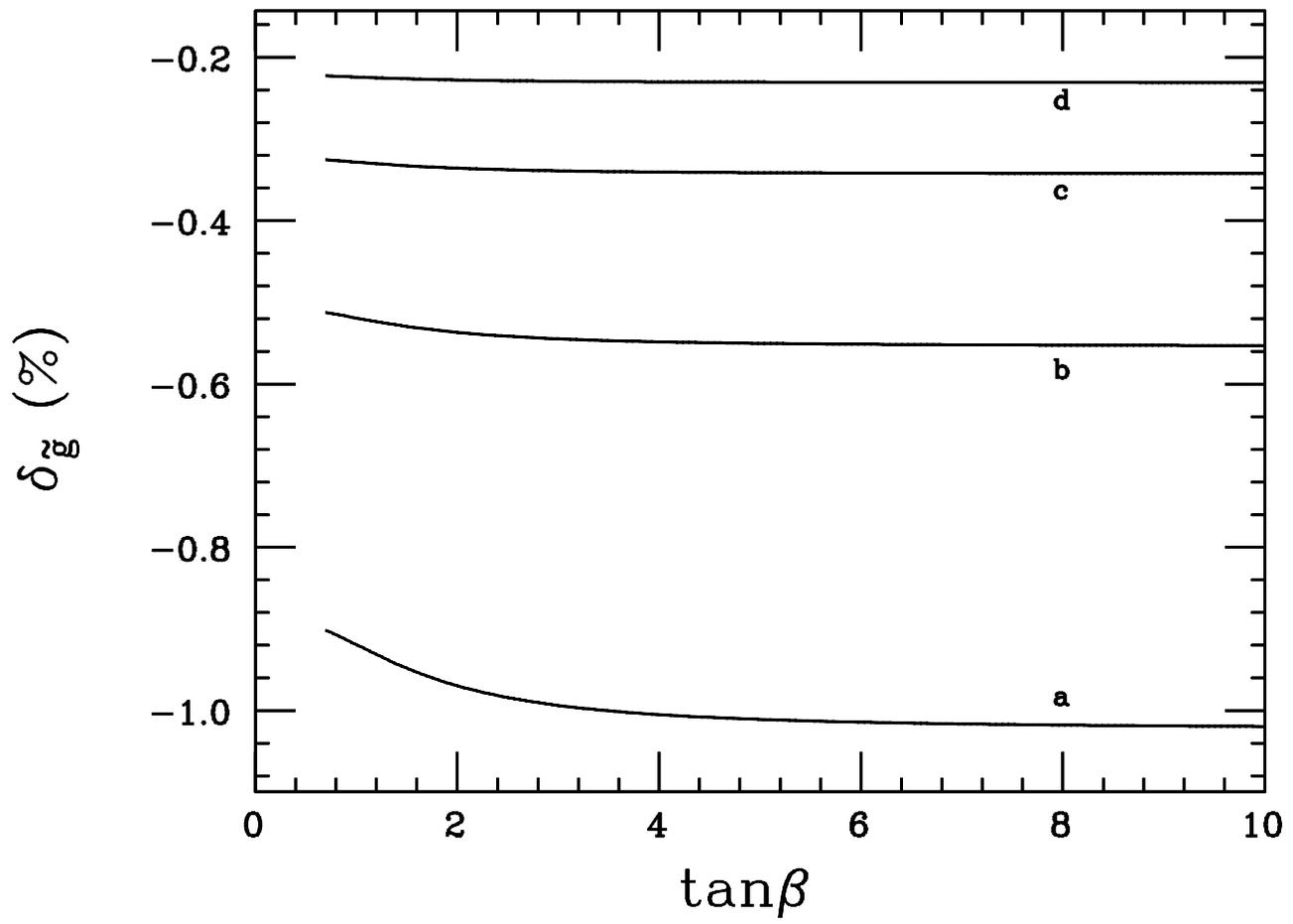
\item{\bf Fig.3} $\delta_{\tilde{g}}$ as a function of $\tan\beta$, for
the same squark and gluino masses as in Fig.2a, and $m_t=174\,GeV$.

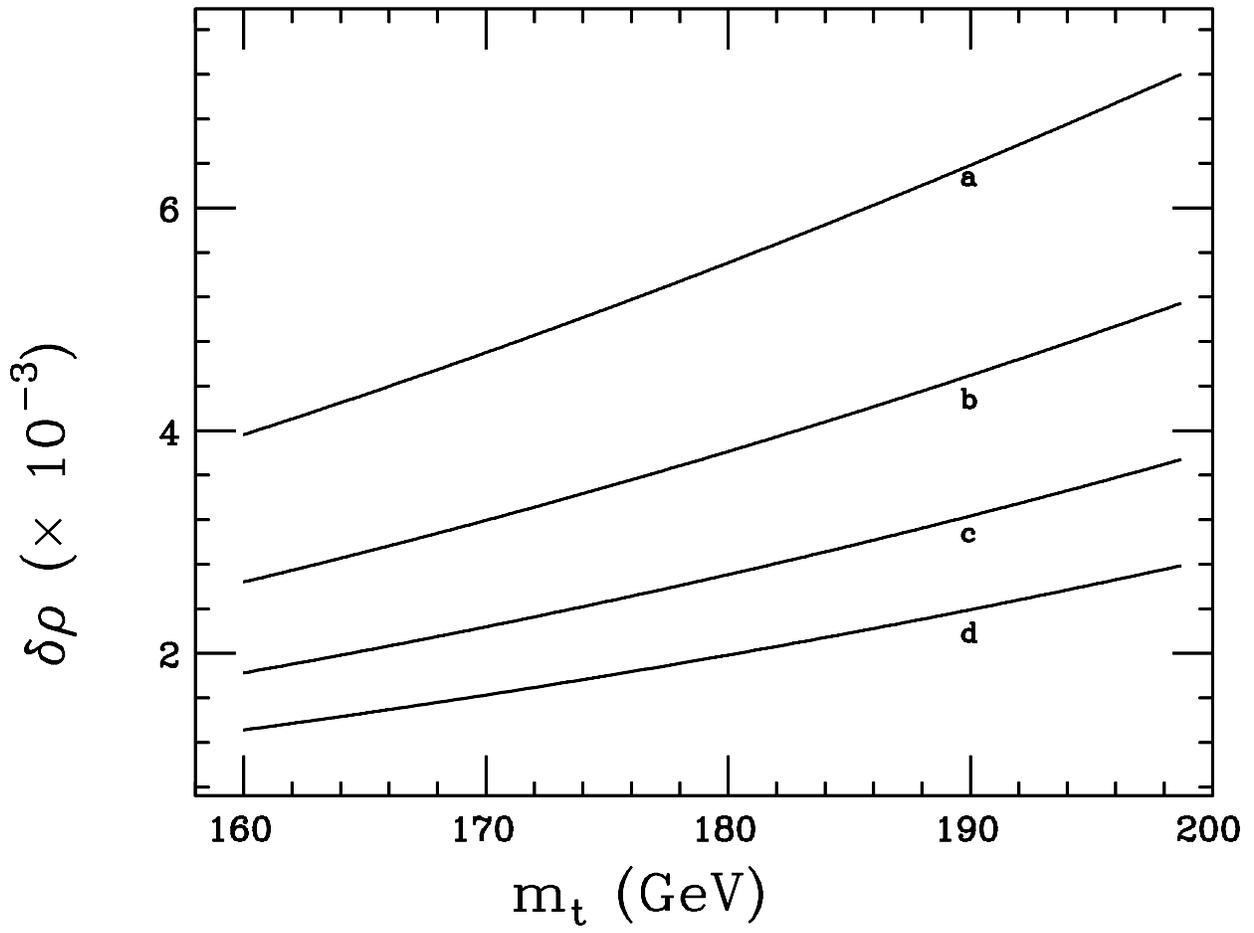
\item{\bf Fig.4} Deviation of the $\rho$-parameter from 1 for the
squark masses used to evaluate $\delta_{\tilde{g}}$ in Fig.2(a).

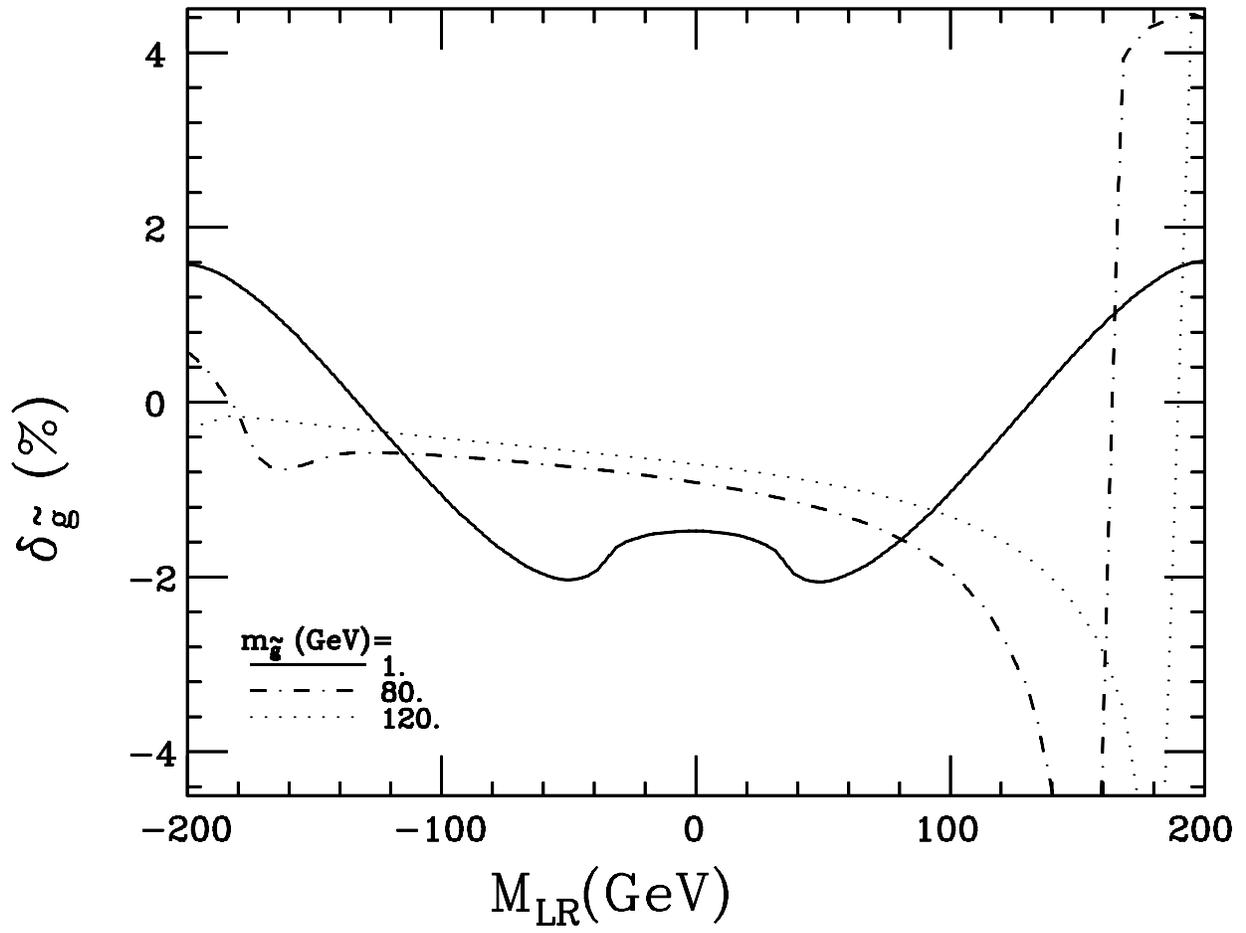
\item{\bf Fig.5} Plot of  $\delta_{\tilde{g}}$ versus the mixing parameter
of the stop mass matrix, $M_{LR}$, for $\theta_t=\pi/4$,
$m_{\tilde{b}}=80\,GeV$ and various
gluino masses ($m_t=174\,GeV$).

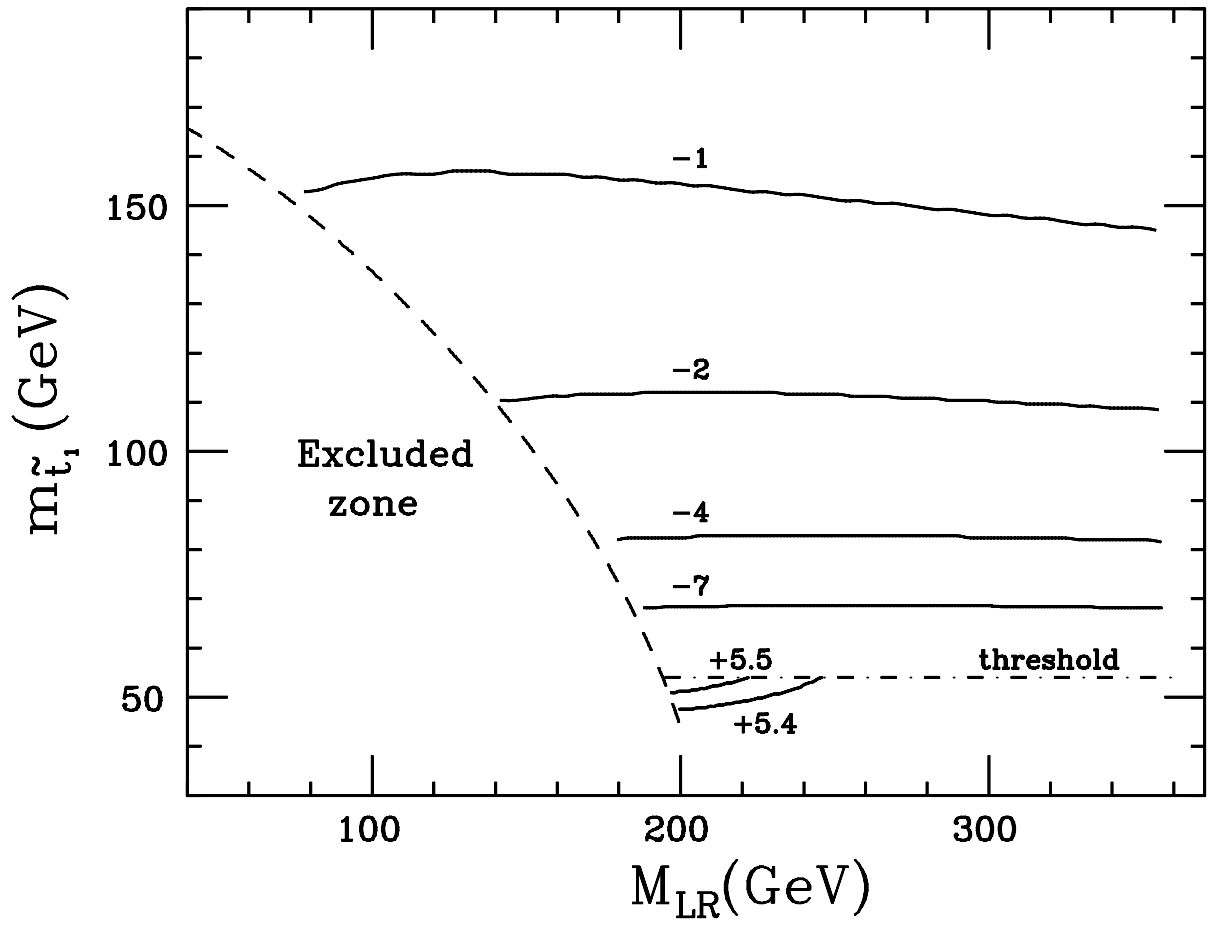
\item{\bf Fig.6} Contour plots of $\delta_{\tilde{g}}$ in the
$(M_{LR}, m_{\tilde{t_1}})$-plane for $m_{\tilde{b}}=m_{\tilde{g}}=120\,GeV$.
The mixing angle $\theta_t$ varies accordingly and the shaded area
is excluded by the condition $M_{\tilde{t}_R}^2>0$ in the stop mass matrix
($m_t=174\,GeV$).

\end{itemize}

\end{document}